\documentclass[preprint,amsfonts,amssymb,amsmath]{revtex4}

\usepackage{times}
\usepackage{graphicx}

\begin{document}

\title{A linear reformulation of the Kuramoto model of self-synchronizing coupled oscillators}

\author{David C. Roberts}  

\affiliation{Theoretical Division and Center for Nonlinear Studies, Los Alamos National Laboratory, Los Alamos, NM} 

\begin{abstract} 
The present paper introduces a linear reformulation of the Kuramoto model describing a self-synchronizing phase transition in a system of globally coupled oscillators that in general have different characteristic frequencies.  The reformulated model provides an alternative coherent framework through which one can analytically tackle synchronization problems that are not amenable to the original Kuramoto analysis.  It allows one to solve explicitly for the synchronization order parameter and the critical point of 1) the full phase-locking transition for a system with a finite number of oscillators (unlike the original Kuramoto model, which is solvable implicitly only in the mean-field limit) and 2) a new class of continuum systems.  It also makes it possible to probe the system's dynamics as it moves towards a steady state.  While discussion in this paper is restricted to systems with global coupling, the new formalism introduced by the linear reformulation also lends itself to solving systems that exhibit local or asymmetric coupling.
\end{abstract}

\date{\today}

\maketitle
\section {Introduction} Spontaneous synchronization of coupled oscillators with different natural frequencies is at the core of many striking phenomena in dynamical systems in realms from biology and physics to social dynamics \cite{syncbook, winfree}.  The complexity of these systems defied attempts to encapsulate them in tractable mathematical formulations until, in 1975, Kuramoto produced a model that he was able to solve exactly for systems containing an infinite number of globally weakly coupled nonlinear oscillators \cite{Kuramoto1, Kuramoto2, crawford, kurreview}.  In such a system, the global coupling strength across the oscillators and the width of the oscillators' initial characteristic frequency distribution determine whether or not the system will self-synchronize, and with these Kuramoto was able to describe a phase transition into a self-synchronizing system.  The Kuramoto model has since provided the basis for many later efforts to explore spontaneous synchronization.  More importantly still, it is a valuable explanatory model in its own right for understanding numerous situations involving synchronization.   In addition to examples in biology (see \cite{syncbook, kurreview} and reference therein) such as neural firing patterns, clouds of fireflies flashing as one, and the coordinated action of cardiac pacemaker cells, many examples of the Kuramoto model have been recently discovered throughout the physical sciences including in such diverse systems as Josephson junction arrays \cite{jjarrays}, collective atomic recoil lasing \cite{voncube},  flavor evolution of oscillating neutrinos \cite{neutrinos}, and phase locking of oscillations in coupled chemical reactions \cite{kiss}.

This paper reformulates the Kuramoto model in terms of linear dynamics, permitting its solution through an eigenvalue/eigenvector approach.  The analysis here is restricted to solving for the critical point of the fully locking transition and the synchronization order parameter beyond this transition; the properties of partially locked states are not considered.  Within this regime, the reformulation of the Kuramoto model has a number of advantages over the original model.  In particular, in the continuum limit, this linear model makes it possible to solve exactly a new class of synchronization models that is distinct from the class of systems for which the Kuramoto model in its original form has an analytic solution.  Furthermore, whereas in the original version of the Kuramoto model the synchronization order parameter is only solvable in the continuum limit and then only implicitly, the alternative form presented below allows the order parameter to be solved explicitly and for any number of oscillators.  In addition, the linearity of the reformulation makes it possible to investigate the time evolution of a system's self-synchronization and lends itself to adaptation to systems that exhibit local and/or asymmetric coupling between oscillators.  

The present paper will begin with an introduction to the original Kuramoto model and its implicit analytic solution in the thermodynamic limit.  Then the linear reformulation and its solution will be presented, and the mapping between the Kuramoto model in its currently accepted form and linear version presented here will be shown.  This will be followed by some examples to demonstrate properties of the linear reformulation.

\section {Kuramoto model} The Kuramoto model \cite{Kuramoto1, Kuramoto2,crawford, kurreview} describes a collection of $N$ oscillators that are weakly coupled:
\begin{equation}
\label{kuramoto}
\dot{\theta}_k=\omega_k+\sum^N_{j \ne k} K_{jk} \sin(\theta_j-\theta_k)
\end{equation}
where $K_{jk}$ is the coupling constant, $\omega_k$ is the characteristic frequency of the $k^{th}$ oscillator, and $\theta_k$ is its phase.  Although Kuramoto's method only yields an analytic solution for a uniform coupling scheme (further discussion below), note that the model's coupling constant is general and does not imply uniform coupling.  Kuramoto was able to exactly solve eq. (\ref{kuramoto}) for a system of globally and uniformly coupled oscillators, i.e. $K_{jk}=K/N$, with the constraints that the system be in the thermodynamic limit, i.e. $N \rightarrow \infty$, and have reached a steady-state; in so doing, he showed the existence of a nontrivial self-synchronization phase transition in such a system.  In this paper, as a measure of the synchronization of a system at a point in time, we will use the amplitude $r$ of the complex phase order parameter,
\begin{equation}
\label{r}
r e^{i \phi} \equiv \frac{1}{N} \sum_{j=1}^N e^{i \theta_j},
\end{equation}
(where $\phi$ represents the collective phase of the synchronized state) which ranges between 0 (no synchronization) and 1 (perfect synchronization); steady-state synchronization of the system will be specified as such.  Kuramoto's result was an implicit equation for $r$ in the synchronized state given by 
\begin{equation}
\label{rkuramoto}
r=Kr \int_{-\frac{\pi}{2}}^{\frac{\pi}{2}} \cos^2 \theta g(K r \sin \theta) d \theta
\end{equation}
where $g=g(\omega_k)$ is the distribution of $\omega_k$ and the frame of reference is the rotating frame in which the frequency of the synchronized solution is zero.  We will refer hereafter to eq. (\ref{rkuramoto}), with the abovementioned accompanying constraints, as the Kuramoto {\em solution}.  The Kuramoto model eq. (\ref{kuramoto}), along with its many variations, has been studied in great detail \cite{kurreview}, and is seen as the standard model for synchronization of globally weakly coupled nonlinear oscillators.  

\section {Linear reformulation and solution} We now propose a linear reformulation of the Kuramoto model of spontaneous synchronization: 
\begin{equation}
\label{model}
\dot{\psi}_k=(i \omega_k-\gamma) \psi_k + \sum_{j\ne k}^N \Omega_{jk}\psi_j,
\end{equation}
where $\Omega_{jk}$ is the coupling constant of this linear model and $\gamma$ is the decay constant to be tuned to bring the amplitude of $\psi_k$ | a complex variable | to a steady state (details given below).  As we will later show, the argument of $\psi_k$ corresponds to $\theta_k$ in eq. (\ref{kuramoto}).  For simplicity, we assume $\gamma>0$, and $\omega_k$ is real.  In this reformulation we consider global coupling because it is convenient and will permit easy comparison of our results with those of previously studied mean-field models, but this approach lends itself equally well to other linear coupling schemes \cite{differences}.

The linear model (eq. \ref{model}) has the simple solution 
\begin{equation}
\vec{\psi}=\sum_{j=1}^N a_j \vec{v}_j e^{\lambda_j t},
\end{equation} 
where $a_j$ are constants determined by the initial conditions, and $\vec{v}_j$ and $\lambda_j$ are the eigenvectors and eigenvalues associated with the matrix defined by the RHS of eq. (\ref{model}).  The synchronizing behavior of the system in the long-time limit is dependent on the eigenvalue(s) with the greatest real part.  Thus in this analysis we will adopt the convention of ordering all eigenvalues by their real part, from $\lambda_1$ (least) to $\lambda_N$ (greatest).  Distinct eigenvalues with the same real part shall arbitrarily be assigned consecutive subscripts within the larger sequence.  In our discussion we will assume $\lambda_N$ is not degenerate.

We tune $\gamma$ so that $\Re[\lambda_N]=0$.  If $\Re[\lambda_{N-1}] \ne 0$ then the solution becomes 
\begin{equation}
\lim_{t \rightarrow \infty} \vec{\psi} = a_N \vec{v}_N e^{i \omega_r t}
\end{equation} 
where the collective frequency of the fully locked state, $\omega_r$, is given by 
\begin{equation}
\omega_r=-i \lambda_N =|\lambda_N|
\end{equation} 
and is related to the collective phase of the locked state by $\phi= \omega_r t$.  As a result, $r$ will tend to a steady-state value between 0 and 1, which (in a finite-$N$ system) indicates full locking, where the entire oscillator population is locked to one particular frequency.  The ensuing analysis will focus on the transition to this fully locked state (the full locking transition).  For finite-$N$ systems, the transition may be from partial locking | where there are subpopulations locked to different frequencies | to full locking, or from incoherence to full locking.  

One can calculate the properties of the steady-state synchronization phase, where only one eigenvector remains in the long-time limit.  To find an expression for $r$ one must determine the eigenvector associated with $\lambda_N$.  Hereafter, for simplicity we assume $\Omega_{jk}=\frac{\Omega}{N}>0$ unless otherwise specified.  The general form of the corresponding eigenvector for $1 \le j \le N$, where $j$ is the index for the components of $\vec{v}_N$, is given by 
\begin{equation}
(v_N)_j=\frac{i (\omega_N-\omega_r) -\Omega/N-\gamma}{i (\omega_j -\omega_r)-\Omega/N -\gamma}.
\end{equation}
The general explicit expression for $r$ of a fully phase-locked system with an arbitrary distribution of $\omega_k$ over finite $N$ in the long-time limit is then
\begin{align}
r&\equiv \left|\frac{1}{N}\sum_{j=1}^N e^{i\theta_j}\right|=\frac{1}{N}\left|\sum_{j=1}^N\frac{\psi_j}{|\psi_j|}\right|=\frac{1}{N}\left| \sum_{j=1}^N \frac{(v_N)_j}{|(v_N)_j|} \right|\notag\\
&= \frac{1}{N}\left| \sum_{j=1}^N  \frac{i (\omega_N-\omega_r) -\Omega/N-\gamma}{i (\omega_j -\omega_r)-\Omega/N -\gamma} \sqrt{\frac{(\omega_j-\omega_r)^2+(\Omega/N+\gamma)^2}{(\omega_N-\omega_r)^2+(\Omega/N+\gamma)^2}} \right|,\label{rgeneral}
\end{align}
which, it should be stressed, is independent of initial conditions.
If we assume the distribution of frequencies to be symmetric about $\omega_r$, i.e. $g(\omega_r - \omega_k)=g(\omega_r +\omega_k)$, then we can simplify eq. (\ref{rgeneral}) to arrive at 
\begin{equation}
\label{rgensimple}
r= \frac{1}{N} \sum_{j=1}^N \left[1+\left( \frac{\omega_j-\omega_r}{\Omega/N+\gamma}\right)^2 \right]^{-1/2}.
\end{equation}
If $r$ goes to a nonzero steady-state value, there is steady-state synchronization of the system.  In a finite system, the full locking transition takes place where $r$ goes between having and not having a steady-state value.  This transition point occurs for a given frequency distribution $g(\omega_k)$ when the following relationship with the critical value of the coupling constant $\Omega_c$ is satisfied: 
\begin{equation}
\Re[\lambda_{N-1}(\Omega_c)] = 0.
\end{equation}

To see the connection between the linear reformulation, eq. (\ref{model}), and the Kuramoto model, eq. (\ref{kuramoto}), we perform the nonlinear transformation $\psi_k(t)=R_k(t)e^{i \theta_k(t)}$ on eq. (\ref{model}) to arrive at 
\begin{equation}
\label{theta}
\dot{\theta}_k=\omega_k+\frac{\Omega}{N} \sum^N_{j \ne k}  \frac{R_j}{R_k} \sin(\theta_j-\theta_k),
\end{equation}
\begin{equation}
\label{Rt}
\dot{R}_k(t)=-\gamma R_k+\frac{\Omega}{N} \sum^N_{j \ne k} R_j  \cos(\theta_j-\theta_k).
\end{equation}

By tuning $\gamma$ so that $\Re(\lambda_N)=0$ we can force each $R_k$ to go to a steady state, namely
\begin{equation}
\lim_{t\to\infty} R_k(t)=|a_N (v_N)_k| = |a_N| \sqrt{\frac{(\omega_N-\omega_r)^2+(\Omega/N+\gamma)^2}{(\omega_k-\omega_r)^2+(\Omega/N+\gamma)^2}}.
\end{equation}
If all $R_k$ go to a steady state for large times, then eq. (\ref{theta}) becomes eq. (\ref{kuramoto}).  Therefore, this linear version maps onto eq. (\ref{kuramoto}) with the effective coupling constant 
\begin{equation}
\tilde K_{jk}=\lim_{t\to\infty} \frac{\Omega}{N} \frac{R_j}{R_k}=\frac{\Omega}{N} \sqrt{\frac{(\omega_k-\omega_r)^2+(\Omega/N+\gamma)^2}{(\omega_j-\omega_r)^2+(\Omega/N+\gamma)^2}}.
\end{equation}
It is important to note that $\tilde K_{jk}$ is independent of initial conditions as $a_N$ cancels out, and that the mapping holds only in the regime in which there is steady-state synchronization (where $\Re[\lambda_{N-1}] \ne 0$).  With this, eq. (\ref{theta}) can be rewritten as 
\begin{equation}
\label{tildetheta}
\dot{\theta}_k=\omega_k+ \sum^N_{j \ne k} \tilde K_{jk} \sin(\theta_j-\theta_k).
\end{equation}
In other words, by introducing the amplitude $R_k$ properly constrained by the decay constant $\gamma$, we would be able to perform a nonlinear transformation of the Kuramoto model eq. (\ref{kuramoto}) (which only has an implicit solution in the infinite-$N$ limit) with an effective coupling constant of $\tilde K_{jk}$ into our linear version eq. (\ref{model}) that can be solved exactly for any $N$.   One can conceive of much more general mappings between eq. (\ref{kuramoto}) and eq. (\ref{model}), such as by allowing each oscillator to have its own independent $\gamma \rightarrow \gamma_k$, or by breaking the assumption of uniform coupling in the linear reformulation, e.g. $\Omega \rightarrow \Omega_{jk}$ (which does not break the linearity) to accommodate a larger class of couplings $K_{jk}$.

\section {Examples} To emphasize the mathematical properties of our linear model, we work through three groups of examples in detail below.  In each, we solve for $r$ using the linear approach and compare the result to the Kuramoto solution, eq. (\ref{rkuramoto}), in the thermodynamic limit (the regime of validity of the Kuramoto solution).

\subsection {Identical oscillators} In our first example we consider a system where all characteristic frequencies are equal, i.e. $\omega_k=\omega$.  This is a simple system which will exhibit steady-state perfect synchronization regardless of initial conditions (as long as $\Omega \ne 0$).  In this case, to solve the linear model (eq. (\ref{model})), we first observe that there are $N-1$ degenerate eigenvalues, each equal to $i\omega-\gamma-\Omega/N$ and one unique eigenvalue, 
\begin{equation}
\lambda_N=i\omega-\gamma+\Omega(N-1)/N, 
\end{equation}
that has an associated eigenvector $v_N=\{1,1,1,...,1\}$.  (Note that the normalization of the eigenvector does not matter for the calculation of the order parameter because $r$ is defined such that every element has a modulus of $1$).  In order for each $| \psi_k |$ to go to a steady state, we set $\gamma=\Omega (N-1)/N $ (consistent with eq. (\ref{Rt}) for a perfectly synchronized state) so that as $t \rightarrow \infty$ all of the degenerate eigenvectors decay away on a time scale $\frac{N}{(N-2)\Omega}$, which for large $N$ reduces to $\sim 1/\Omega$.  Assuming $\Omega \ne 0$, the solution to the reformulation with $\omega_k=\omega$ is then 
\begin{equation}
\lim_{t \rightarrow \infty} \vec{\psi} = a_N \vec{v}_N e^{i \omega t}.
\end{equation}

The synchronization order parameter can now be computed easily, giving the steady-state value of $r=1$ for long times.  Because each element in the sum is normalized by its magnitude, the long-time steady-state behavior of $r$ is independent of the initial conditions represented by $a_j$.  Since $\tilde K_{jk} \rightarrow \Omega/N$ in the long-time limit in this example, in the infinite-$N$ limit the linear model's solution, eq. (\ref{rgensimple}), and the Kuramoto solution, eq. (\ref{rkuramoto}), give the identical result of $r=1$ assuming the frequency distribution $\omega_k=\omega$. It is also instructive to take small perturbations around the characteristic frequencies, i.e. $\omega_k=\omega+\epsilon \eta_k$ where $0 < \epsilon \ll 1$ and $\eta_k$ are symmetrically distributed around $0$.   In this case, to leading order 
\begin{equation}
r \approx 1-\epsilon^2 \frac{\Delta^2}{2 \Omega^2},
\end{equation} 
where $\Delta^2$ is the variance given by $\Delta^2=\frac{1}{N} \sum_{j=1}^N \eta_j^2$. 

\subsection {Bimodal distribution of characteristic frequencies} For our second example, we look at a system in which both steady-state partial synchronization and a full-locking transition occur.   Consider a bimodal distribution where $\omega_j=\omega_0 - \frac{\Delta}{2}$ for $1 \le j \le N/2$ and $\omega_j=\omega_0 + \frac{\Delta}{2}$ for $N/2 < j \le N$  where $\Delta \ge 0$ is the width of the distribution (we will assume $N$ is even for simplicity).  The solution to this problem maps onto a system of two coupled oscillators because of the pecularity of the distribution. 

Solving the linear reformulation, we see that there are $(N-2)/2$ degenerate eigenvalues given by $i (\omega_0 - \frac{\Delta}{2})-\frac{\Omega}{N}-\gamma$ and $(N-2)/2$ degenerate eigenvalues given by $i (\omega_0 + \frac{\Delta}{2}) -\frac{\Omega}{N}-\gamma$.  The final two eigenvalues $\lambda_N$ and $\lambda_{N-1}$ are given by
\begin{equation}
\lambda_{N,N-1}=i \omega_0 -\gamma+ \frac{\Omega(N-2)}{2 N} \pm \frac{\sqrt{\Omega^2-\Delta^2}}{2}
\end{equation}
where the `$+$' refers to $\lambda_N$, and the `$-$' refers to $\lambda_{N-1}$.  

First let us take the situation where $\Omega < \Delta$.  Here we shall see that $r$ does not go to a steady-state value and therefore that the system will not fully lock and no steady-state synchronization occurs.  Unlike the previous example, there are not one but two distinct eigenvalues with the largest real part, namely 
$\Re[\lambda_{N-1}] = \Re[\lambda_N]=-\gamma+ \frac{\Omega(N-2)}{2 N}.$
So even after setting $\gamma=\frac{\Omega(N-2)}{2 N}$ we find that
\begin{equation}
\lim_{t \rightarrow \infty} \vec{\psi} = a_{(N-1)} \vec{v}_{(N-1)} e^{i \left( \omega_0 - \frac{\sqrt{\Delta^2-\Omega^2}}{2} \right) t}+ a_{N} \vec{v}_{N} e^{i \left( \omega_0 + \frac{\sqrt{\Delta^2-\Omega^2}}{2} \right) t}.
\end{equation} 
It is clear that as long as $\Omega < \Delta$, $r$ will never go to a steady-state value and hence the whole system will never fully lock or reach steady-state synchronization.  However, independent of all initial conditions and assuming $\Omega>0$, the population of oscillators will split into two perfectly synchronized subsets with different collective frequencies given by $\Im[\lambda_{N,N-1}] = \omega_0 \pm \frac{\sqrt{\Delta^2-\Omega^2}}{2}$.

Now let us consider the situation where $\Omega>\Delta$.  If we set $\gamma= \frac{\Omega(N-2)}{2 N} + \frac{\sqrt{\Omega^2-\Delta^2}}{2}$ (note $\gamma \rightarrow \Omega(N-1)/N$ for large $\Omega$, consistent with eq. (\ref{Rt}) for a perfectly synchronized state) then, on a time scale given by $1/ \sqrt{\Omega^2-\Delta^2}$, all but one eigenvalue die out over time:  
\begin{equation}
\lim_{t \rightarrow \infty} \vec{\psi} = a_N \vec{v}_N e^{i \omega_0 t}
\end{equation}
where the components of the eigenvector are given by 
\begin{equation}
(v_N)_j=\frac{i \Delta - 2 \Omega/N- \sqrt{\Omega^2-\Delta^2}}{-i \Delta - 2 \Omega/N-  \sqrt{\Omega^2-\Delta^2}}
\end{equation}
for $1 \le j \le N/2$ and $(v_N)_j=1$ for $N/2 < j \le N$.  Thus $r$ goes to a steady-state value:
\begin{equation}
\label{rfinal}
r= \sqrt{\frac{1+\sqrt{1-\left( \frac{\Delta}{\Omega} \right)^2}}{2}} 
\end{equation}
for $\Omega>\Delta$.  It is important to note that although this equation is true for finite $N$ | unlike the Kuramoto solution eq. (\ref{rkuramoto}) | the expression for $r$ is independent of $N$.  

Since when $\Omega<\Delta$ the system does not fully lock and no steady-state synchronization occurs, and when $\Omega>\Delta$ there is full locking and steady-state synchronization, it is clear that when $\Omega_c = \Delta$ a ``first-order'' full-locking transition occurs, marked by steady-state partial synchronization beyond $r_c=1/\sqrt{2}$.  Using eq. (\ref{rfinal}) we can deduce a square-root scaling law of the order parameter near the full-locking transition (which in this case is also the transition into a steady-state self-synchronizing system), i.e.
\begin{equation}
r-r_c \approx \frac{1}{2}\sqrt{\frac{\Omega-\Omega_c}{\Omega}}.
\end{equation}

As in the previous example, this bimodal example has the property of $\tilde K_{jk} \rightarrow \Omega/N$ in the long time limit,  so the linear form maps onto the Kuramoto model (eq. (\ref{kuramoto})) with the simple transformation $\Omega \rightarrow K$.  From the above discussion therefore, if one assumes a bimodal frequency distribution, i.e. $g(\omega)=\frac{\delta(\omega-\Delta/2)}{2}+\frac{\delta(\omega+\Delta/2)}{2}$, then solving eq. (\ref{rkuramoto}), which is only valid in the infinite-$N$ limit, gives our result, eq. (\ref{rfinal}).

\subsection {Continuum limit} In this final example, we work exclusively in the continuum limit and demonstrate that, even within this regime where the Kuramoto solution is valid, the linear reformulation still has the advantage of a simpler solution for the order parameter and opens up a new class of exactly solvable synchronization models.  We first determine the critical point of the locking transition and the synchronization order
parameter given by the linear model for a system of oscillators with a general symmetric characteristic frequency distribution and a particular effective coupling.  Systems with such coupling, which we will consider below, go from completely incoherent to fully locked at one point in parameter space, i.e. there are no partially locked states; the partial-locking and full-locking transitions merge \cite{samept}. Next, we solve for the critical point and the synchronization order parameter for a Lorentzian frequency distribution and a uniform frequency distribution.  We then look at how the result differs from the traditional uniform coupling solution.

We begin by solving for $r$, $\gamma$, and the locking transition point (also the synchronization critical point), each in terms of $\Omega$ and a general symmetric frequency distribution.  As $N \rightarrow \infty$, eq. (\ref{rgensimple}) becomes
\begin{equation}
\label{rint}
r=\int_{-\infty}^{\infty} d \omega g(\omega)  \left[1+\left( \frac{\omega-\omega_r}{\gamma}\right)^2 \right]^{-1/2},
\end{equation} 
which holds for $\Omega>\Omega_c$.  The anomalous scaling properties of this solution are discussed in \cite{razvan}.  The linear reformulation eq. (\ref{model}) in the continuum limit has the same form as the equation describing the evolution of the fundamental mode in the stability analysis in \cite{StrogMiro} of the incoherent state in the Kuramoto model:
\begin{equation}
\label{lo}
\dot \psi(\omega,t)=(i \omega - \gamma) \psi(\omega,t) +\Omega \int_{-\infty}^{\infty} g( \omega') \psi(\omega',t) d \omega',
\end{equation}
which has an effective coupling in the continuum Kuramoto model of 
\begin{equation}
\label{effcoup}
K(\omega,\omega')= \Omega \sqrt{\frac{(\omega' - \omega_r)^2 + \gamma^2}{(\omega - \omega_r)^2 + \gamma^2}}.
\end{equation}
Following the analysis in \cite{StrogMiro}, the spectrum of the linear operator of the RHS of eq. (\ref{lo}) comprises an eigenvalue $\lambda_N$ at the origin of the complex plane and a continuous line of eigenvalues along $-\gamma$ of the real axis.  If $\Omega > \Omega_c$, $\gamma>0$ and $\lambda_N$ lies apart from the other eigenvalues, so there is full locking and synchronization in the long-time limit.  However if $\Omega \le \Omega_c$, then $\gamma=0$, $\lambda_N$ merges with the continuum along the imaginary axis, and contribution from every frequency remains in the long-time limit, which results in $r=0$ as all oscillators are ``drifting''.  This implies a second-order phase transition.  In this case, the partial-locking transition (often referred to as the synchronization phase transition) and the full-locking transition occur at the same point.  Furthermore, the coupling described by eq. (\ref{effcoup}) is such that, at the onset of the transition, there is a uniform distribution of phases from $0$ to $2 \pi$ across the oscillator population (i.e. $r \to 0$ as $\Omega \to \Omega_c$).  We can determine $\gamma$ using the self-consistency equation \cite{StrogMiro} 
\begin{equation}
\label{selfcon}
1 = \Omega \int_{-\infty}^{\infty} \frac{g(\omega)}{\gamma - i \omega} d\omega.
\end{equation}
At most one solution for $\gamma$ exists and $\gamma \ge 0$ \cite{strogatz}.  Therefore the system will reach steady-state synchronization for $\gamma > 0$, and the critical point $\Omega_c$ occurs as $\gamma \rightarrow 0^+$.  This can easily be shown to be 
\begin{equation}
\label{omegac}
\Omega_c= \frac{1}{\pi g(0)}.
\end{equation}

Now, considering a Lorentzian distribution of frequencies about $\omega_r$, i.e. $g(\omega-\omega_r)=\frac{\Delta}{\pi [\Delta^2+(\omega-\omega_r)^2]}$, from eq. (\ref{omegac}) we obtain $\Omega_c = \Delta$.  Integrating eq. (\ref{selfcon}) gives $\gamma_{lor} = \Omega - \Delta$.  So where $\Omega > \Omega_c$, 
\begin{equation}
r_{lor}=\frac{2}{\pi} \cos^{-1} \left( \frac{\Omega_c}{\Omega-\Omega_c} \right) \left[ 1-\left( \frac{\Omega_c}{\Omega-\Omega_c} \right)^2\right]^{-1/2}
\end{equation}
in the long-time limit.  Similarly for a uniform distribution of frequencies about $\omega_r$, i.e. $g(\omega-\omega_r)=\frac{1}{\pi \Delta} $ for $|\omega-\omega_r|<\pi \Delta/2$ and $0$ otherwise, the same steps give us $\Omega_c=\Delta$, $\gamma_{unif}=\frac{\Delta \pi}{2} \cot(\frac{\pi \Delta}{2 \Omega})$, and 
\begin{equation}
r_{unif}=\cot \left( \frac{\pi}{2} \frac{\Omega_c}{ \Omega} \right) \sinh^{-1} \left[ \tan \left( \frac{\pi}{2} \frac{\Omega_c}{\Omega  }\right)\right]
\end{equation}
in the steady state for $\Omega > \Omega_c$.  (It is interesting to note that $r_{unif} \ge r_{lor}$ for the entire parameter regime.)  By contrast, the result one obtains from the Kuramoto solution eq. (\ref{rkuramoto}) for a Lorentzian distribution of frequencies (where a partial population of drifting oscillators can occur) is $r=\sqrt{1-2 \Delta/K}$ \cite{Kuramoto1,Kuramoto2} for $K \ge K_c = 2\Delta$ and $r=0$ otherwise; and for a uniform distribution of frequencies one cannot find an explicit solution.  

\section {Closing comments} In closing, two points deserve mention.  First, although the dynamics of the linear model are in principle completely solvable, they are not the same as the dynamics of the original Kuramoto model since the mapping between eq. (\ref{model}) and eq. (\ref{kuramoto}) only formally holds when all $R_j$ go to a steady state.  Second, while in this paper we have restricted ourselves only to global coupling, the same analysis should be applicable to any linear coupling scheme, including local or asymmetric, as long as one is able to determine the largest eigenvalues and eigenvectors of the RHS of eq. (\ref{model}).\\  

{\bf Acknowledgements:}
The author is grateful for informative discussions with Eli Ben-Naim, Razvan Teodorescu, and Colm Connaughton.  The author would also like to thank Steven Strogatz and Renato Mirollo for their careful reading of the manuscript and their suggestions for clarifying the language of the paper.


\begin{thebibliography}{9}
\bibitem{differences} The synchronization analysis of amplitude oscillators in this paper differs from analyses that incorporate cubic nonlinearities (see for example \cite{strogatz} and \cite{strogatz2}) in that here phase, not amplitude, synchronization is considered.  Whereas in these other models the amplitude order parameter $\rho \equiv \frac{1}{N} \left| \sum_{j=1}^N \psi_j \right| =  \frac{1}{N} \left| \sum_{j=1}^N R_j
e^{i \theta_j} \right|$ has an upper bound of 1 and is used as a measure of amplitude synchronization, in the present model $\rho$ has no upper bound and the amplitude is not a free variable.  Furthermore,
in the linear model presented in this paper the amplitude decays for systems of uncoupled oscillators, unlike in the models with cubic nonlinearities.
\bibitem{samept}   Uniformly coupled continuum systems | those that have an analytic Kuramoto solution eq. (\ref{rkuramoto}) | with uniform characteristic frequency distributions also exhibit this absence of partially locked states \cite{MiroStrog}.
\end{thebibliography}
\end{document}